\documentstyle[12pt,epsf]{article}

\newcommand{\be}{\begin{equation}}
\newcommand{\ee}{\end{equation}}
\newcommand{\ben}{\begin{eqnarray}\displaystyle}
\newcommand{\een}{\end{eqnarray}}
\newcommand{\refb}[1]{(\ref{#1})}
\newcommand{\p}{\partial}
\newcommand{\sectiono}[1]{\section{#1}\setcounter{equation}{0}}

\newcommand{\II}{{\cal I}}
\newcommand{\CC}{{\cal C}}
\newcommand{\SS}{{\cal S}}
\newcommand{\VV}{{\cal V}}
\newcommand{\HH}{{\cal H}}

\newcommand{\NN}{{\cal N}}

\newcommand{\KK}{{\cal K}}
\newcommand{\QQ}{{\cal Q}}

\newcommand{\wt}{\widetilde}
\newcommand{\wh}{\widehat}

\newcommand{\bmu}{{\bar\mu}}
\newcommand{\bnu}{{\bar\nu}}
\newcommand{\xl}{{x^L}}
\newcommand{\xr}{{ x^R}}

\begin{document}
{}~
\hfill\vbox{\hbox{hep-th/0105058}\hbox{CTP-MIT-3132}
\hbox{PUPT-1985} 
\hbox{NSF-ITP-01-35}
}\break

\vskip .6cm

\centerline{\large \bf  Half-strings, Projectors, and Multiple D-branes}

\medskip

\centerline{\large \bf  in Vacuum String Field Theory }

\vspace*{4.0ex}

\centerline{\large \rm Leonardo Rastelli$^a$, Ashoke Sen$^b$ and Barton
Zwiebach$^c$}

\vspace*{4.0ex}

\centerline{\large \it ~$^a$Department of Physics }

\centerline{\large \it Princeton University, Princeton, NJ 08540, 
USA}  

\centerline{E-mail:
        rastelli@feynman.princeton.edu}

\vspace*{2ex}

\centerline{\large \it ~$^b$Harish-Chandra Research
Institute}

\centerline{\large \it  Chhatnag Road, Jhusi,
Allahabad 211019, INDIA}

\centerline {and}
\centerline{\large \it Institute for Theoretical Physics}

\centerline{\large \it 
University of California, Santa Barbara, CA 93106, USA
}

\centerline{E-mail: asen@thwgs.cern.ch, sen@mri.ernet.in}

\vspace*{2ex}

\centerline{\large \it $^c$Center for Theoretical Physics}

\centerline{\large \it
Massachussetts Institute of Technology,}

\centerline{\large \it Cambridge,
MA 02139, USA}

\centerline{E-mail: zwiebach@mitlns.mit.edu}

\vspace*{5.0ex}

\centerline{\bf Abstract}
\bigskip

A sliver state is a classical solution of the string field theory
of the tachyon vacuum that represents a  background with a single
D25-brane. We show that  the sliver wavefunctional factors into 
functionals of the left and right halves of the string, and hence 
can be naturally regarded as  a rank-one projector in a space of 
half-string functionals. By developing an algebraic oscillator 
approach we are able construct higher rank projectors that describe 
configurations of multiple D-branes of various dimensionalities 
and located at arbitrary positions.  The results bear remarkable
similarities with  non-commutative solitons.

\vfill \eject

\baselineskip=16pt

\tableofcontents

\newpage

\sectiono{Introduction and summary}  \label{s1} 

Vacuum string field theory is a new approach to open string field
theory that uses directly the mysterious open string tachyon vacuum
for the description of the theory \cite{0012251,0102112}.
Among all possible open string backgrounds the tachyon vacuum is
particularly natural given its physically expected uniqueness as the
endpoint of all processes of tachyon condensation. It is also
becoming clear that this vacuum string field theory is structurally
much simpler than conventional cubic open string 
field theory \cite{WITTENBSFT}.\footnote{String field actions with
purely ghost kinetic operators arose previously in ref.\cite{HORO}.}
In fact, under the consistent assumption of ghost/matter factorization,
equations of motion in the matter sector are simply equations for a
projector. A closed form expression for a particular projector  has
already been obtained using a very special string field called the
``sliver"  \cite{0006240,0008252,0102112}.   This sliver state represents
a classical
solution describing a space-filling D25 brane. 
Classical solutions representing (single) 
D-branes of lower dimensions have
also been constructed in \cite{0102112} by using the algebraic
description of the sliver as a starting point for generalization.  
In another work \cite{rsz} we have shown that the 
universal 
definition of the sliver allows one to
construct the solution representing the D-brane associated to a deformed
boundary conformal field theory.

In this paper we extend the above results to construct multiple D-brane
solutions of various dimensions situated at 
various positions.   
Conceptually, the starting point for this analysis is the realization that
the sliver can be viewed as a rank-one projector in a certain state
space. As emphasized in \cite{rsz}, the geometrical
picture of the sliver indicates that it is a state originating from
a Riemann surface where the left-half and the right-half of the open 
string are ``as
far as they can be'' from each other. Thus, one can entertain the
possibility that the sliver functional 
factors into the product of a functionals
of the left-half  and  the 
right half of the open string. 
We test explicitly this factorization  using the representation 
of the sliver in flat 26 dimensional
space-time as a squeezed state in the oscillator basis \cite{0008252},
{\it i.e.},  an
exponential of an oscillator bilinear  acting on the Fock vacuum.
By rewriting this expression in terms of the coordinates $x^L$ of the
left-half of
the string and the coordinates $x^R$ of the 
right-half of the string we
confirm
that the
sliver wave-functional factors into a product of the form $f(x^L)
f(x^R)$. 
Regarding string fields 
as matrices with row index labeled by $x^L$
and column index labeled by $x^R$, the 
$*$-product becomes  matrix product, and the sliver
is recognized as a rank-one projection operator in a space
of half-string functionals. 

This makes it clear
how to construct higher rank projectors. For example, in order to get a
rank two
projector, we construct a new half-string functional $g$ that is
orthogonal to $f$,
and simply take the sum $f(x^L)f(x^R)+g(x^L)g(x^R)$.  Since
$f(x^L)f(x^R)$ and  
$g(x^L) g(x^R)$ are each rank-one projectors, 
and they project onto orthogonal
subspaces, their sum is a rank-two projector. This
is readily generalized to build rank-$N$ projectors.
 While single D-brane solutions can be 
identified as rank-one
projection operators, a solution representing $N$ 
D-branes corresponds to a
projection operator of rank $N$. 
This situation is analogous to
the one that arises in the study of non-commutative
solitons \cite{0003160,0005006,0005031,0010034,0010060}.
We note in passing that in contrast to the
sliver, the ``identity'' string field $\II$, 
which acts as the identity of
the $*$-product, represents a projector into the full half-string
 Hilbert
space and thus, in some sense, represents a configuration of infinite
number of D-branes.

\smallskip
The half-string approach to open string field theory, arising
from observations 
in Witten's original paper 
\cite{WITTENBSFT}, was developed 
in \cite{comma, BORDES, BORDES2}, 
and the possible relevance of this formalism to
the sliver was anticipated in \cite{0008252}.  
The straightforward splitting of the string coordinates into those
of the left-half  and the right-half of the string 
holds only for
zero momentum string states. For string fields carrying momentum there is
additional dependence on the zero mode of the coordinate or equivalently
the string mid-point $x^M$, 
which cannot be thought of as belonging to
either
the left-half or the right-half of the string. Thus the
above method is not directly applicable to generalize the construction
of a D-$p$-brane solution ($p<25$) \cite{0102112} into
multiple D-$p$-brane solutions. 
However we show that  it is possible to construct projectors by
taking the string field of the form $f(x^L;x^M)
f(x^R;x^M)$ with a certain 
functional form for $f$.
It is very likely that the lower dimensional D-brane states constructed
in ref.\cite{0102112}
are of this type.
We can then generalize this to construct higher rank projectors following
the same principle as before,
but we do not pursue this line of argument any further.

Motivated 
by the somewhat formal nature of 
functional manipulations, and by the fact that our evidence for 
left/right factorization of the sliver is only numerical, we develop
a self-contained algebraic approach that incorporates all
the desired features in a completely unambiguous way.  Going
back to the D-25-brane case, we use the oscillator basis 
to build left/right
projectors. These projectors can be directly generalized to
lower dimensional brane solutions, 
with no need of referring to the functional integral half-string
language, 
because the 
(infinite dimensional) matrices that appear in the construction of 
the D-25-brane solution and lower dimensional D-brane solutions have
very similar
structure~\cite{0008252}. 
This allows us to construct
multiple D-brane solutions of lower dimensions. Finally, we can also
generalize this method to construct multiple
D-brane solutions of various dimensions, situated at 
various positions. 

\medskip
The paper is organized as follows. In section \ref{sr1} we give a
brief review of the results of refs.\cite{0012251,0102112}. In section 
\ref{s2} we
review the construction of the sliver state as an exponential of
bilinears in matter oscillators acting
on the Fock vacuum. We also list various identities
satisfied by the infinite
dimensional matrices appearing in the construction of the sliver, and
discuss how to compute $*$-product of states built on the sliver by the
action of matter oscillators. These
identities play a crucial role in the later verification that we have
constructed higher rank projectors.

Section \ref{swhp} is devoted to the functional approach based
on half-strings.  
As in \cite{BORDES} we give the transformations
between the Fourier modes of the full string coordinates and those of
the left
and right-half of the string. This is used in section \ref{smds2} to
write the sliver
functional in terms of the Fourier modes of the half-strings, and to show
numerically that it factorizes into a product $f(x^L)f(x^R)$ with $f$ a
gaussian. 
We construct projectors $g(x^L)g(x^R)$ orthogonal to the sliver by taking
$g$ orthogonal to $f$, and show how to build higher rank projectors. 
We also discuss briefly how to  
generalize this procedure to D-branes of lower dimension, but do not
study this in detail.

The intuition developed in section \ref{swhp} allows us to give in
section 
\ref{smds} a self-contained rigorous 
treatment within the algebraic oscillator
approach.  We construct higher rank projectors in section \ref{smds3}
and confirm
explicitly that their superposition satisfies the
string equations of motion and has the right tension to describe
multiple D-brane states. Section \ref{smdx} contains a short discussion on
the issue of Lorentz invariance of our solutions, and also the origin of
the Chan-Paton factors.  
In section \ref{smds4}
we construct multiple
D-$p$-brane solutions for $p<25$. As pointed out before, while the idea of
left/right factorization of the functional 
could be pursued for lower branes as well along the lines of
section 4.4,  
there is no need for this 
since the algebraic
analysis of section
\ref{smds3} can be generalized in a straightforward manner. 
Finally in
sections \ref{smdy} and \ref{smds5} we combine all the techniques
developed earlier to construct multiple D-brane solutions with D-branes of
various dimensions and situated at 
various positions. A brief discussion of the results is given in
section~\ref{s10}.

We conclude this introduction with the note that results similar to the
ones discussed in this paper have been obtained independently 
by Gross and Taylor
\cite{gross-taylor}.

\sectiono{Review of vacuum string field theory} 
\label{sr1}

In this section we shall briefly describe the results of
refs.\cite{0012251,0102112}.  
In these papers we proposed a
form of
the string field theory action around the open bosonic string
tachyon vacuum 
and discussed classical solutions 
describing D-branes of various dimensions. 
In order to write concretely this theory (which is
formally manifestly background independent \cite{rsz}) we choose
to use the state space $\HH$ of the combined matter-ghost boundary
conformal field theory (BCFT) describing the D25-brane.
The string
field
$\Psi$ is a state of ghost number one  
in $\HH$  and the
string field action is given by:
\be \label{eo1}
\SS (\Psi) \equiv \,-\, {1\over g_0^2}\,\,\bigg[\, {1\over 2} \langle
\,\Psi \, ,
 \, \QQ\, \Psi
\rangle + {1\over 3}\langle \,\Psi \, , \, \Psi *
\Psi \rangle \bigg] \,,
\ee
where $g_0$ is the open
string coupling constant, $\QQ$ is an operator made purely of ghost
fields, $\langle \, ,  \, \rangle$
denotes the BPZ inner product,
and $*$ denotes the usual $*$-product of the string
fields \cite{WITTENBSFT}. 
$\QQ$ satisfies the requirements:
\ben \label{eFINp}
&& \QQ^2 = 0, \nonumber \\
&& \QQ (A * B) = (\QQ A) * B + (-1)^{A} A * (\QQ B)\, , \\
&& \langle \, \QQ A , B \,\rangle = - (-)^A \langle A , \QQ B \rangle
\,. \nonumber
\een 
The action \refb{eo1} is then invariant under the gauge transformation:
\be \label{egtrs}
\delta\Psi = \QQ\Lambda + \Psi * \Lambda - \Lambda * \Psi \, ,
\ee
for any ghost number zero state $\Lambda$ in $\HH$. Ref.\cite{0012251}
contains candidate operators $\QQ$ satisfying these constraints; for our
analysis we shall not need to make a specific choice of $\QQ$.

The equations of motion are
\be \label{eo2}
\QQ \Psi + \Psi * \Psi = 0\, .
\ee
In ref.\cite{0102112} we made the ansatz that all D-$p$-brane solutions in
this theory have the factorized form:
\be \label{eo3}
\Psi = \Psi_g \otimes \Psi_m\, ,
\ee
where $\Psi_g$
denotes a state obtained by acting with the ghost
oscillators on the SL(2,R) invariant vacuum of the ghost BCFT, and
$\Psi_m$  is a
state obtained by acting with matter operators on the SL(2,R)
invariant
vacuum of the matter BCFT.
Let us denote by
$*^g$ and $*^m$ the star product in the ghost and matter sector
respectively.
Eq.\refb{eo2} then factorizes as
\be \label{eo4}
\QQ \Psi_g = - \Psi_g *^g \Psi_g \,,
\ee
and
\be \label{eo5}
\Psi_m = \Psi_m *^m \Psi_m\, .
\ee
We further assumed that the ghost part $\Psi_g$ is universal for all
D-$p$-brane solutions. Under this assumption the ratio of energies
associated with two different D-brane solutions, with matter parts
$\Psi_m'$ and $\Psi_m$ respectively, is given by:
\be \label{eo7}
{\langle \Psi_m' | \Psi_m'\rangle_m \over \langle \Psi_m |
\Psi_m\rangle_m} \, ,
\ee
with $\langle | \rangle_m$ denoting BPZ inner product in the matter BCFT.
Thus the ghost part drops out of this calculation.

In ref.\cite{0102112} we constructed the matter part of the solution for
different D-$p$-branes, and verified that we get the correct ratio of
tensions of D-$p$-branes using eq.\refb{eo7}. The matter part of the
D-25-brane solution was given by the sliver state $|\Xi\rangle$ which will
play a central role in the analysis of this paper. 
The algebraic construction 
of $|\Xi\rangle$ will be reviewed in section \ref{s2} (a detailed
geometrical discussion of the  sliver can be found
in \cite{rsz}).  

{}In the rest of the paper we shall deal with the matter part of the
string state. In order to avoid cluttering up the various formul\ae\ we
shall find it convenient to drop the subscript and the superscript $m$
from various states, inner products and $*$-products; but it should be
understood that all operations refer to the matter sector only.

\sectiono{The oscillator representation of the sliver} \label{s2}

\newcommand{\lo}{\xi}
\newcommand{\gl}{z}
\newcommand{\gf}{\wh w}

In this section we discuss the oscillator representation of the sliver.
Section \ref{s3.2} contains a description of the sliver and some useful
identities involving the infinite dimensional matrices that arise in this
description. Section \ref{a2} contains some results on $*$-products of
states built on the sliver which will be useful for later analysis.

\subsection{Description of the sliver}  \label{s3.2}

Here we wish to review the construction of the matter part of the
sliver in the oscillator representation and consider the basic algebraic
properties that guarantee that the 
multiplication of two
slivers gives a sliver. In fact we follow the discussion
of Kostelecky and Potting \cite{0008252} who gave the
first algebraic construction of a state that would star
multiply to itself in the matter sector. This discussion
was simplified in \cite{0102112} where due attention was
also paid to normalization factors that guarantee that
the states satisfy precisely the projector 
equation \refb{eo5}.  
Here we summarize the relevant parts of this analysis
which will be used in the present paper. In doing so we
also identify some 
infinite dimensional matrices with the properties of projection operators.
These matrices will be useful in the construction of multiple D-brane
solutions in section \ref{smds}.

\medskip
In order to star multiply two states $|A\rangle$ and $|B\rangle$
we must calculate
\be \label{esstar}
|A\rangle*|B\rangle_3 =  {}_1\langle A| {}_2\langle B| V_3\rangle_{123} \,,
\ee
where $|~\rangle_r$ denotes a state in the $r$-th string Hilbert
space, and $|~\rangle_{123}$ denotes a state in the product of the Hilbert
space of three strings. The key ingredient here is the three-string vertex
$|V_3\rangle_{123}$.  While the vertex has nontrivial
momentum dependence, if the states $A$ and $B$ are at zero 
momentum, the star product gives a zero momentum state that can be
calculated using 
\be \label{e9}
|V_3 \rangle_{123} = 
\exp \Bigl(-  {1\over 2} \sum_{{r,s}}
  a^{(r)\dagger}\cdot V^{rs}
\cdot a^{(s)\dagger}\Bigr)
|0\rangle_{123} \,,
\ee
and the rule $\langle 0 | 0 \rangle =1$. Here the $V^{rs}$,
with $r,s= 1,2,3$, are infinite matrices $V^{rs}_{mn}$ ($m,n=
1,\cdots \infty$) satisfying the cyclicity condition
$V^{rs} = V^{r+1,s+1}$ and the symmetry condition
$(V^{rs})^T = V^{sr}$.  
These properties imply that out of
the nine matrices, three: $V^{11}, V^{12}$ and $V^{21}$, can
be used to obtain all others. 
$a^{(r)\mu\dagger}_m$ ($0\le\mu\le 25$) 
denote oscillators in the $r$-th string Hilbert space. 
For simplicity, the Lorentz and the 
oscillator indices, and the Minkowski matrix
$\eta_{\mu\nu}$
used to contract the Lorentz indices,  
have all been suppressed in eq.\refb{e9}. We shall follow
this convention throughout the paper.

One now introduces  
\be \label{ea1}
M^{rs} \equiv CV^{rs}\,, \quad \hbox{with}\quad
 C_{mn} = (-1)^m \delta_{mn},
\quad m,n\geq 1 \, .
\ee
These matrices can be shown to satisfy the following
properties:
\ben \label{ea222}
&& CV^{rs} = V^{sr} C\,, \quad 
(V^{rs})^T = V^{sr}\, , \nonumber \\ 
&& (M^{rs})^T =
M^{rs}\, , \quad CM^{rs} C = M^{sr}\,, \quad [M^{rs}, M^{r's'} ] = 0\, . 
\een
In particular note that all the $M$ matrices commute with each other.
Defining $X \equiv M^{11}$, the three relevant matrices are
$X, M^{12}$ and $M^{21}$. Explicit formulae exist that allow
their explicit 
computation \cite{gross-jevicki,gj2,cremmer,samuel}.\footnote{Our 
convention \refb{esstar} for describing the star product differs
slightly from that of ref.\cite{gross-jevicki}, the net effect of which
is that the explicit expression for the matrix $V^{rs}$
listed in appendix A of
ref.\cite{0102112}, version 1,  
is actually the expression for the matrix
$V^{sr}$.
Since all the explicit computations performed of ref.\cite{0102112}
involved $V^{sr}$ and $V^{rs}$ symmetrically, this does not affect any
of the calculations in that paper.}
 
They can be
shown in general grounds to satisfy the following useful relations:
\ben \label{ea33}
&& X+ M^{12} + M^{21} =1 \, , \nonumber \\
&& M^{12} M^{21} = X^2 -X\, , \nonumber \\
&& (M^{12})^2 + (M^{21})^2 = 1- X^2\, , \nonumber \\
&& (M^{12})^3 + (M^{21})^3 = 2X^3 - 3 X^2 + 1 = (1-X)^2
(1+2X)\,,
\een
which gives
\be \label{ea33a}
(M^{12}- M^{21})^2 = (1- X)(1+ 3X)  \, .
\ee
The state in the matter Hilbert space  
that multiplies to itself turns out to take
the form \cite{0008252,0102112}
\be \label{e21a}
|\Psi\rangle = \NN^{26}    
\exp \Bigl(-{1\over 2}\, a^\dagger \cdot S\cdot a^\dagger
\Bigr) |0\rangle\,, \quad \NN= \{\det(1 - X)\det (1+T) \}^{1/2}\,, \quad
S = CT
\ee
where the matrix $T$  satisfies $CTC= T$ and the
equation
\be \label{ea44}
XT^2 - (1+X) T + X=0\, ,
\ee
which gives
\be \label{ea44a}
T = (2 X)^{-1} ( 1 + X -
\sqrt{(1+3 X) (1 - X)}). 
\ee
In taking the square root we pick that branch which,
for small $X$, goes as $(1+X)$.

In \cite{0102112} we identified this state as the sliver by
computing numerically the matrix $S$ using the equation
above and comparing the state obtained this way with the matter part of
the sliver $|\Xi\rangle$, which 
using the techniques of ref.\cite{LPP},   
can be expressed as 
\be \label{e21aa}
|\Xi\rangle = \wh\NN^{26}  
\exp \Bigl(-{1\over 2}\, a^\dagger \cdot \wh S\cdot a^\dagger
\Bigr) |0\rangle\,, 
\ee
\be \label{es1}
\wh S_{mn} = -{1\over \sqrt{mn}} \,  \ointop
{d w\over
2\pi i}\, \ointop {dz\over 2\pi i} {1 \over z^n\, w^m }
{df(z)\over dz} {df(w)\over dw} {1\over  (f(z) - f(w))^2}\, ,
\ee
with
$f(z)= \tan^{-1}(z)$. We found close agreement between the numerical
values
of $S_{mn}$ and exact answers for $\wh S_{mn}$.
This gave convincing evidence that $|\Psi\rangle = |\Xi\rangle$.

\medskip
We record, in passing, the following additional
relations between $T$ and $X$ following from the form
of the solution and the equation that relates them (\refb{ea44},
\refb{ea44a}): 
\be 
\label{ehg}
{1-TX\over 1-X} = {1\over 1-T} \,,\quad 
{1-T\over 1+ T } = \sqrt{1 - X\over 1+ 3X}\,, \quad 
{X\over 1-X} ={ T\over (1-T)^2}\,.
\ee

\medskip
We conclude this subsection by constructing a pair
of projectors that will be very useful in explicit
computations of star products.
We define the matrices 
\ben
\label{proj}
\rho_1 &&= {1\over (1+T)
(1-X)}\Bigl[ M^{12} (1-TX)  + T (M^{21})^2  \Bigr] \,, \\
\rho_2 && =  {1\over (1+T)
(1-X)} \Bigl[ M^{21} (1-TX) + T (M^{12})^2\Bigr] \,. \nonumber
\een 
One readily verifies that they satisfy the following properties:
\be
\rho_1^T = \rho_1\,, \quad \rho_2^T = \rho_2\,, \quad C\rho_1 C = \rho_2\,,
\ee
and more importantly
\ben
\label{proj2}
 \rho_1 + \rho_2 &&= 1 \,, \\
 \rho_1 - \rho_2 &&=  {M^{12} - M^{21}\over \sqrt{(1-X) (1+3 X)}} \, . 
\nonumber
\een
{}From eq.\refb{ea33a} we see 
that the square of the second right hand side is
the unit matrix. 
Thus $(\rho_1-
\rho_2)^2 = 1$, and this together with the squared version of the first
equation gives
\be 
\rho_1 \rho_2 = 0 \,.
\ee
This equation is also easily verified directly.
Multiplying the first equation in \refb{proj2} 
by $\rho_1$ and
alternatively 
by $\rho_2$ we get
\be  \label{erhoproj}  
\rho_1 \rho_1 = \rho_1\,, \quad \rho_2 \rho_2 = \rho_2\,.
\ee
This shows that $\rho_1$ and $\rho_2$ are projection operators into
orthogonal subspaces, and the $C$ exchanges these two subspaces.
We will see later on that  $\rho_1$ and $\rho_2$ project into
the oscillators 
of the right and the left  
half of the  
string
respectively. Finally we record the identities
\ben \label{ea8}
M^{21} \rho_1  + M^{12} \rho_2 &&= X(T-1)  \\
(M^{12})^2 \rho_1  + (M^{21})^2 \rho_2 &&= (1-TX)^2  \,. \nonumber
\een

\subsection{Computing $*$-products of states} \label{a2}

In this subsection we will give the methods needed 
to compute the star product of states involving the (matter)
sliver and matter oscillators acting on the sliver.
The methods are equally applicable (we give
an example) to the case where the products
involve a factor where matter oscillators
act on the identity string field.
 
As seen earlier,
the matter part of the sliver state is given by
\be \label{eb1}
|\Xi\rangle = \NN^{26} \exp\Big(-{1\over 2} 
a^{\dagger}
\cdot S \cdot a^{\dagger}\Big)|0\rangle\, .
\ee
Coherent states are defined by letting exponentials
of the creation operator act on the vacuum. Treating
the sliver as the vacuum we introduce coherent like
states of the form
\be \label{eb2}
|\Xi_\beta \rangle =
\exp \Bigl( \sum_{n=1}^\infty
(-)^{n+1}\beta_{\mu n} a_n^{\mu \dagger}\Bigr) |\Xi
\rangle =
\exp ( -a^\dagger \cdot C\beta ) |\Xi
\rangle\, .
\ee
As built, the states satisfy a simple BPZ conjugation
property:
\be \label{eb3}
\langle \Xi_\beta| =  \langle \Xi |
\exp \Bigl( \sum_{n=1}^\infty \beta_{n\mu} a_n^\mu
\Bigr) = \langle \Xi | \exp ( \beta \cdot a)\, .
\ee
We  compute the $*$ product of two such states using the 
procedure 
discussed in refs.\cite{0008252, 0102112}. We begin by 
writing out the product using two by two matrices encoding
the oscillators of strings one and two:
\ben  
\Bigl( |\Xi_{\beta_1}\rangle * |\Xi_{\beta_2}
\rangle \Bigr)_{(3)}  
&=& \Bigl(\exp ( -a^\dagger \cdot
C\beta_1) |\Xi
\rangle * \exp ( -a^\dagger \cdot
C\beta_2 ) |\Xi\rangle\Bigr)_{(3)} \nonumber \\
&=& {}_{(1)}\langle \Xi| \exp (\beta_1\cdot a_{(1)}) \,   
 \,\, {}_{(2)}\langle \Xi| \exp (\beta_2\cdot a_{(2)})
  |V_{123}\rangle \nonumber \\
&=&\langle 0_{12} |
\exp\Bigl(  \beta \cdot a -{1\over 2} a\cdot \Sigma
\cdot a  \Bigr) 
\exp\Bigl( -{1\over 2} a^\dagger\cdot {\cal V} \cdot
a^\dagger - \chi^T \cdot a^\dagger\Bigr) 
|0_{12}\rangle\nonumber \\
&& \qquad \cdot
\exp\Bigl( - {1\over 2} a_{(3)}^\dagger \cdot V^{11} 
\cdot a_{(3)}^\dagger\Bigr) |0_3\rangle \,\,,
\een
where  $a = (a_{(1)}, a_{(2)})$, and 
\be \label{erepeat1}
\Sigma = \pmatrix{S&0\cr 0&S} , \quad {\cal V} =
\pmatrix{V^{11}& V^{12} \cr V^{21} & V^{22} }, \quad 
\beta = \pmatrix{\beta_1 \cr \beta_2} , \quad
\chi^T = ( a_{(3)}^\dagger V^{12}\,,\, 
a_{(3)}^\dagger  V^{21})  \, . 
\ee 
Explicit evaluation continues by using the equation
\ben
\label{key}
&& \langle 0 | \exp\Big(\beta_i a_i -{1\over 2} P_{ij} a_i
a_j\Big) \exp\Big(-\chi_i a^\dagger_i -{1\over 2} Q_{ij} a^\dagger_i
a^\dagger_j\Big) |0\rangle \\
&&\hskip-20pt  =   \det(K)^{-1/2} \exp\Big(-\chi^T\, K^{-1} \,\beta
-{1\over 2}
\beta^T \,Q \, K^{-1} \beta - {1\over 2}\chi^T\, K^{-1}P\chi\Big)\,
,\,\, K\equiv 1- PQ\,.  \nonumber
\een
At this time we realize that since $|\Xi\rangle * |\Xi\rangle =
|\Xi\rangle$ the result of the product is a sliver with 
exponentials acting on it; the exponentials that contain
$\beta$. This gives 
\be \label{eb4}
|\Xi_{\beta_1}\rangle * |\Xi_{\beta_2}\rangle =
\exp \Bigl( -\chi^T \, \KK^{-1} \, \beta -
{1\over 2}\, \beta^T \, {\cal V} \KK^{-1} \, \beta
\Bigr) |\Xi\rangle\, ,\quad  \KK = (1 - \Sigma\VV)\, .
\ee
The expression for $\KK^{-1}$, needed above is simple
to obtain given that all the relevant submatrices commute.
One finds that 
\be \label{submat}
\KK^{-1} =(1-
\Sigma{\cal V})^{-1} = {1\over (1+T) (1-X)} 
\pmatrix{1-TX & TM^{12}\cr TM^{21} & 1-TX}  \,.
\ee
We now recognize that the projectors $\rho_1$ and $\rho_2$
defined in \refb{proj} make an appearance in the oscillator
term of \refb{eb4}
\ben
\label{enterproj}
-\chi^T \KK^{-1} \beta
&=& -a^\dagger
\cdot C \, (M^{12}, M^{21} ) \KK^{-1}\beta = 
- a^\dagger \cdot C \,(\rho_1, \rho_2) \beta  \\
&=& - a^\dagger C \,
\cdot (\rho_1 \, \beta_1+ 
\rho_2 \, \beta_2) \,. \nonumber
\een
We can verify that 
\ben \label{eb7}
{\cal C} (\beta_1, \beta_2) &&\equiv 
{1\over 2}\, (\beta_1, \beta_2) \, {\cal V} \KK^{-1}
\, \pmatrix{\beta_1 \cr \beta_2}   \\
&&={1\over 2}\, (\beta_1, \beta_2){1\over (1+T)(1-X)} 
\pmatrix{V^{11}(1-T) & V^{12} \cr V^{21} & V^{11}(1-T) }
\pmatrix{\beta_1 \cr \beta_2}  \,. \nonumber
\een
Since the matrix in between is symmetric we have
\be \label{eb8}
{\cal C} (\beta_1, \beta_2) = {\cal C} (\beta_2, \beta_1) \, .
\ee
Using \refb{enterproj} and \refb{eb7} we finally have:
\be \label{eb9}
|\Xi_{\beta_1}\rangle * |\Xi_{\beta_2}\rangle 
= \exp \Bigl(  -
{\cal C} (\beta_1, \beta_2) \Bigr)  |\Xi_{\rho_1\beta_1
+\rho_2\beta_2}\rangle \, .
\ee
This is a useful relation that allows one to 
compute $*$-products  
of slivers acted by oscillators by simple differentiation.
In particular, 
using eq.\refb{eb2} we get
\ben \label{eb11}
 (a^{\mu_1\dagger}_{m_1}\cdots a^{\mu_k\dagger}_{m_k} |\Xi\rangle) * 
(a^{\nu_1\dagger}_{n_1}\cdots a^{\nu_l\dagger}_{n_l} |\Xi\rangle) 
&=&
(-1)^{\sum_{i=1}^k (m_i + 1) + \sum_{j=1}^l (n_j + 1)} \\
&& \hskip-65pt \Bigl( {\partial \over \partial \beta_{1m_1\mu_1}} \cdots
{\partial
\over
\partial
\beta_{1m_k\mu_k}} {\partial \over \partial \beta_{2n_1\nu_1}}  
\cdots {\partial \over \partial
\beta_{2n_l\nu_l}}\big(|\Xi_{\beta_1}\rangle *
|\Xi_{\beta_2}\rangle\big)\Bigr)_{\beta_1=\beta_2=0}\, .
\nonumber 
\een

\medskip
\noindent
Since
$\rho_1+\rho_2=1$, for
$\beta_1=\beta_2$ eq.\refb{eb9} reduces to
\be \label{eb10}
|\Xi_{\beta}\rangle * |\Xi_{\beta}\rangle 
= \exp \Bigl(  -
{\cal C} (\beta, \beta) \Bigr)  |\Xi_\beta\rangle\, .
\ee
Using the definition of $\CC$ in \refb{eb7} and  
equations \refb{ea44} and \refb{ehg} one can show that
$\CC (\beta, \beta)$ simplifies down to 
\be
\label{sympcc}
\CC (\beta , \beta) =
{1\over 2}\,
\beta C (1-T)^{-1}\beta\,.
\ee
It follows from \refb{eb10} that by adjusting the
normalization of the $\Xi_\beta$ state  
\be
\label{pbeta}
P_\beta \equiv \exp \Bigl(  
{\cal C} (\beta, \beta) \Bigr)  |\Xi_\beta\rangle \,,
\ee
we obtain projectors  
\be
\label{eproj}
P_\beta * P_\beta = P_\beta\,.
\ee
Using eq.\refb{key} one can also check that
\be \label{enormpb}
\langle P_\beta | P_\beta\rangle = \langle\Xi | \Xi\rangle\, .
\ee

\subsection{The identity string field and BPZ inner products}\label{trid}

\noindent
The same methods can be used to multiply coherent states on
the sliver times coherent states on the  identity string
field.\footnote{For some remarkable properties of the 
identity field, see ref.\cite{0105024}.} If we restrict
ourselves to the matter sector
the identity takes the form 
\be \label{eident}
|\II\rangle = [\det (1-X)]^{13} 
                 \exp (-{1\over 2} 
a^\dagger \cdot C \cdot a^\dagger)\, ,  
\ee
where the determinant prefactor has been
included to guarantee that $\II * \II = \II$.
One can also explicitly verify that
\be
\label{idsliv}
\II * \Xi = \Xi * \II = \Xi\, , 
\ee
and, more generally,
\be
\label{idslivg}
\II * \Xi_\beta = \Xi_\beta * \II = \Xi_\beta \, .
\ee
This allows us to
simplify the computation of the BPZ norm of any projector 
$P$ satisfying both $P*\II = \II*P =P$ and $P*P=P$ with  
exact unit normalization.
For this we have
\be
\label{bpzid}
\langle P | P \rangle = \langle \II *P | P \rangle
= \langle \II | P*P\rangle = \langle \II | P \rangle\,.
\ee
where we have used the property $\langle A*B, C\rangle = \langle
A , B*C\rangle$ of the full BPZ product restricted to the matter
BPZ product $\langle \cdot | \cdot \rangle$. 
The above relation is not formal, we have verified directly 
that $\langle \II |\Xi \rangle  = \langle \Xi | \Xi\rangle$ and
that $\langle \II |P_\beta \rangle  = \langle P_\beta | P_\beta\rangle
=\langle \Xi | \Xi\rangle$, as seen in \refb{enormpb}.

In open string field theory one defines a trace as Tr$(\Psi) =
\langle \II|\Psi\rangle$, this is sometimes called 
 `integration', and denoted as $\int \Psi$.  In this notation
we would have that equation \refb{bpzid} reads
Tr$(P) = \langle P | P \rangle$. While the sliver will be interpreted
as a rank one projector, and thus one would expect Tr$(\Xi) =1$, the
conformal anomaly in the matter BCFT does not allow us to have both
$\Xi*\Xi = \Xi$ and $\langle \Xi | \Xi\rangle =1$. We normalized
the matter sliver so that the projection condition is satisfied 
exactly. In this case, however, $\langle \Xi | \Xi\rangle$ does not
equal one,  
and in fact is numerically seen to vanish 
in the infinite level limit. 
We expect that when we fully understand the ghost sector of the theory
and calculate the value of the string field theory action for a D-brane
solution, we shall get a finite answer for the action.

\medskip
In order to consider star products involving the identity
and the sliver, both with oscillators acting on them, we introduce
\be
\label{iind}
|{\cal I}_\beta\rangle = 
\exp \Bigl( \sum_{n=1}^\infty
(-)^{n+1}\beta_n a_n^\dagger\Bigr) |{\cal I}
\rangle =
\exp ( -a^\dagger \cdot C\beta ) |{\cal I}
\rangle \,. 
\ee
This time we find
\ben
|\Xi_\alpha\rangle * |{\cal I}_\beta\rangle 
 &=& 
\exp \Bigl( - {1\over 2}  \beta C \,{X(1-T)\over (1-X)} \,\beta - 
\alpha C {M^{12} \over 1-X} \beta \Bigr)
|\Xi_{\alpha + (1+T) \rho_2 \beta}\rangle \nonumber\\
|{\cal I}_\beta\rangle * |\Xi_\alpha\rangle  
 &=& 
\exp \Bigl( - {1\over 2}  \beta C \,{X(1-T)\over (1-X)} \,\beta 
 - \alpha C {M^{21} \over 1-X} \beta\Bigr) 
|\Xi_{\alpha + (1+T) \rho_1 \beta}\rangle \,. 
\een 

\sectiono{Sliver wavefunctional, half-strings and projectors} 
\label{swhp}

\newcommand{\sm}{*^m}

In this section we shall examine the representation of
string fields as functionals of half strings. This viewpoint
is possible at least for the case of zero momentum string fields.
It leads to the realization that the sliver functional 
factors into functionals of the left and right halves of the string,
allowing its interpretation as a rank-one projector in the
space of half-string functionals. We construct higher rank
projectors -- these are 
solution of the equations of motion \refb{eo5} representing multiple
D25-branes.

\subsection{Zero momentum string field as a matrix} \label{smds1}

The string field equation in the matter sector is given by 
\be \label{eg1}  
\Psi * \Psi = \Psi\, .
\ee
Thus if we can regard the string field as an operator acting on
some
vector space where $*$ has the interpretation of product of operators,
then $\Psi_m$ is a
projection operator in this vector space. Furthermore, in analogy with
the results in
non-commutative solitons \cite{0003160}, 
we expect that in
order to describe a single
D-brane, $\Psi_m$ should be a projection operator into a single
state in this vector space.

A possible operator interpretation of the string field was suggested in
Witten's original paper \cite{WITTENBSFT}, and was further developed in
ref.\cite{comma, BORDES, BORDES2}. 
In this picture the string field is viewed  as a
matrix where the role of the row index and the
column index are taken by the left-half and the 
right-half of the string respectively.
In order
to make this more concrete, let us consider 
the standard mode expansion of
the open string coordinate \cite{gross-jevicki}: 
\be \label{eg2}
X^\mu(\sigma) = x^\mu_0 + \sqrt 2 \sum_{n=1}^\infty x^\mu_n
\cos(n\sigma)\,,\quad \hbox{for}\quad 0\le\sigma\le
\pi\, . 
\ee
Now let us introduce coordinates
$X^{L\mu}$
and $X^{R\mu}$ for the left and the right half of the string as 
follows\footnote{Here the half strings are parameterized
both from $\sigma=0$ to $\sigma=\pi$, as opposed to
the parameterization of \cite{BORDES} where the half strings
are parameterized from $0$ to $\pi/2$.
}:
\ben \label{eg3}
X^{L\mu}(\sigma) &=& X^\mu(\sigma/2) - 
\,\,X^\mu(\pi/2)\,\,, \,\,\qquad \hbox{for}\quad 0\le\sigma\le
\pi\, . \nonumber\\
X^{R\mu}(\sigma) &=&
X^\mu(\pi - \sigma/2) - X^\mu(\pi/2), 
\qquad \hskip-9pt\hbox{for}\quad 0\le\sigma\le
\pi\, .
\een
$X^{L \mu}(\sigma)$ and $X^{R \mu}(\sigma)$ satisfy the usual Neumann
boundary condition at $\sigma=0$ and a Dirichlet boundary condition at
$\sigma=\pi$. Thus they have expansions of the form:
\ben \label{eg4}
X^{L\mu}(\sigma) &=& \sqrt 2 \sum_{n=1}^\infty x^{L\mu}_n
\cos((n-{1\over 2})\sigma),  \\
 X^{R\mu}(\sigma) &=& \sqrt 2
\sum_{n=1}^\infty x^{R\mu}_n
\cos((n-{1\over   
2})\sigma)\, .\nonumber
\een
Comparing \refb{eg2} and \refb{eg4} we get an expression for
the full open string modes in terms of the modes of the left-half
and the modes of the right-half:
\be \label{eg5}
x^\mu_n = A_{nm}^+ \,x^{L\mu}_m +  A^-_{nm} \, x^{R\mu}_m, \qquad
m,n\ge 1\, ,
\ee
where the matrices $A^\pm$ are  
\be \label{eg7}
A_{nm}^\pm =\pm{1\over 2} \delta_{n, 2m-1} +{1\over 2\pi} 
\epsilon (n,m) \Big({1\over 2 m + n - 1} +
{1\over 2 m - n - 1}\Big) \, ,
\ee
and 
\be   \epsilon (n,m) =
(1 + (-1)^n) (-1)^{m+{1\over 2} n - 1} \,.
\ee
Alternatively we can write the left-half modes and right
half modes in terms of the full string modes
\ben \label{eg6}
x^{L\mu}_m &=& \wt A_{mn}^+ \,x^\mu_n, \\
 x^{R\mu}_m &=& \wt
A_{mn}^-\, x^\mu_n,
\qquad m,n\ge 1\, , \nonumber
\een
where one finds
\be \label{eg8}
\wt A_{mn}^\pm  =2 A_{nm}^\pm  -{1\over \pi}\, \epsilon(n,m)\,
\Big( {2\over 2 m -1}\Big) \, . 
\ee
Note that
\be \label{egtwist}
A^+ = C A^-, \quad \wt A^+ = \wt A^- C\, ,
\ee
where $C_{mn}=(-1)^{n} \delta_{mn}$ is the twist operator.
Note also that the relationship between $x_n^\mu$ and $(x_n^{L\mu}$,
$x_n^{R\mu})$ does not involve the zero mode $x_0^\mu$ of $X^\mu$. 
Finally, we observe 
 that by their definition, the various matrices
must satisfy the relations
\ben
A^+ \wt A^+ + A^- \wt A^- = 1\,, && \\ 
\wt A^\pm A^\mp = 0\,,  && \nonumber \\
\wt A^\pm A^\pm = 1\, . && \nonumber
\een
By the last equation, the  $\wt A^\pm$ matrices are
left inverses of the $A^\pm$ matrices, but not  right
inverses, as it follows from the first of the equations.

A general string field configuration can be regarded as a functional of
$X^\mu(\sigma)$, or equivalently a function of the infinite set of
coordinates $x^\mu_n$. Now suppose we have a translationally invariant
string field configuration. In this case it is independent of $x_0^\mu$
and we can regard this as a function 
$\psi(\{x^{L\mu}_n\}, \{x^{R\mu}_n\})$ of the
collection of modes of the left and the right half of the string. 
(The sliver is an example of such a state). 
We will use vector notation to represent these collections
of modes
\ben    
&&\xl \equiv  = 
\{ x^{L\mu}_n \, | \, n=1, \cdots \infty; \, \mu = 0, \cdots
25\} \,, \nonumber \\
&&\xr \equiv  = 
\{ x^{R\mu}_n \, | \, n=1, \cdots \infty; \, \mu = 0, \cdots
25\} \,.
\een
We can also
regard the function
$\psi(\xl, \xr)$ 
as an
infinite dimensional matrix, with the row index labeled by the
modes in 
$\xl$ and the column index labeled by the modes in $\xr$. 
The reality condition on the string
field is the hermiticity of this matrix:
\be \label{eherm}
\psi^*(\xl, \xr)= \psi(\xr, \xl)\, ,
\ee
where the $^*$ as a superscript denotes complex conjugation.
Twist symmetry, on the other hand,
exchanges the left and right
half-strings, so it acts as transposition of the matrix:
twist even (odd) string
fields correspond to symmetric (antisymmetric)
matrices in half-string space.
Furthermore, given two such
functions
$\psi(\xl, \xr)$ and $\chi(\xl, \xr)$, their
$*$ product is given by \cite{WITTENBSFT} 
\be \label{eg9}
(\psi*\chi) \,(\xl, \xr) = 
\int 
[\hbox{d} y]\,\,
\psi(\xl,  y) \,\chi ( y, \xr)\, .
\ee
Thus in this notation
the $*$-product becomes a generalized 
 matrix product. It is clear
that the vector space on which these matrices act is the space
of functionals of the half-string coordinates $\xl$ (or
$\xr$). A projection operator $P$ into a one dimensional subspace of
the half string Hilbert space, spanned by some appropriately normalized
functional $f$, will correspond to a functional of the form:
\be \label{eg10}
\psi_P(\xl,
\xr) = f(\xl) f^*(\xr)\, .
\ee
The two factors in this expression are related by conjugation
in order to satisfy condition \refb{eherm}.
The condition $\psi_P * \psi_P = \psi_P$ requires that
\be
\label{normc}
\int [\hbox{d}y]  f^*(y)  f(y) = 1 \,.  
\ee
By the formal properties of the original 
open string field theory 
construction
one has  $\langle A, B\rangle = \int  A*B$ where $\int $ has the
interpretation of a trace, namely identification of the left
and right halves of the string,
together 
with an integration over 
the string-midpoint coordinate $x^{M \, \mu}=X^\mu(\pi/2)$. 
Applied to a projector
$P$ with associated wavefunction $\psi_P (x^L, x^R)
= f^*(x^L) f(x^R)$, and focusing only in the matter sector
    we would find 
\ben
   \langle P, P \rangle &=& \int P*P = \int P = V 
\int [\hbox{d}x^L][\hbox{d}x^R] \delta (x^L
- x^R)
    \psi_P (x^L, x^R) 
\\   
& =& V\int [\hbox{d}y] f^*(y) f(y)   
= V\, ,
   \nonumber  \een
where $V$ is the space-time volume coming from integration over 
the string midpoint $x^{M \, \mu}$. 
   This shows that (formally) rank-one 
projectors are expected to
    have BPZ normalization $V$. 
In our case, due to conformal anomalies,
    while the matter sliver squares precisely as a projector, its BPZ
    norm approaches zero as the level is increased~\cite{0102112}. 
The above argument applies
    to string fields at zero momentum, thus the alternate projector
    constructed 
in ref.\cite{0102112} representing lower dimensional D-branes
need not have the same BPZ norm as the sliver. Note that 
the above discussion is the functional counterpart of
the algebraic  discussion in subsection \ref{trid}.

\subsection{The left-right factorization 
of the sliver wavefunctional} \label{smds2}

The first question we would like to address is: is the sliver a
projection operator into a one dimensional subspace in the sense just
described? In order to answer this question we need to express the sliver
wave-function as a function of $x^{L\mu}_n$, $x^{R\mu}_n$ and then see if
it
factorizes in the sense of eq.\refb{eg10}. We start with the definition of
the sliver written in the harmonic oscillator basis:
\be \label{eg11}
|\Xi\rangle = \NN^{26} \exp(-{1\over 2} \eta_{\mu\nu}\,\,
a_m^{\mu\dagger}\,  S_{mn}\,
a_n^{\nu\dagger})
|0\rangle\, .
\ee
We now note 
the relation between the operators\footnote{We
shall
denote by $\hat
x^\mu_n$ the coordinate operators and by $x_n^\mu$ the eigenvalues of
these operators.} $\hat x_n^\mu$ and
$a_m^{\mu\dagger}$, $a_m^\mu$: 
\be \label{eg12}
\hat x_n^\mu = {i\over 2} \,
\sqrt{2\over n}\, (a_n^\mu - a_n^{\mu\dagger})\,,\quad
\to \quad  \hat x = {i\over 2} \, E \cdot ( a - a^\dagger ) \,, \quad
E_{nm} = \delta _{nm} \sqrt{{2\over n}} \,,
\ee
which we also write in compact matrix notation using the
matrix $E$ defined above. Using this 
one can express
the position eigenstate in the harmonic oscillator basis:
\be \label{eg13}
\langle 
\vec x | = K_0^{26} \, \langle 0|
\exp\Bigl(-x \cdot E^{-2} \cdot x 
+  2 i \,a\cdot E^{-1} \cdot x + {1\over 2} a\cdot
a\Bigr)\, ,
\ee
where $K_0$ is a normalization constant whose value we shall not need to
know.
Thus the sliver wave-function, expressed in the position basis, is given
by
\be \label{eg14}
\psi_\Xi(\vec x 
) = \langle \vec x \, 
| \Xi\rangle
= \wt\NN^{26} \exp(-{1\over 2}\, x\cdot V \cdot x)\, .
\ee
The evaluation of this contraction is done with the general
formula \refb{key} in section \ref{a2}. One finds that the normalization
factor is $\wt\NN = \NN \, K_0(\det(1+S))^{-1/2}$ 
and that the matrix $V$
above is given as 
\be \label{eg15}
V = 2 E^{-2} - 4  
E^{-1} S (1+S)^{-1} E^{-1}\quad\to \quad  V_{mn} = n
\delta_{mn} - 2\sqrt{mn} (S(1+S)^{-1})_{mn}\, .
\ee
We can now rewrite $\psi_\Xi$ as a function of $x^L$ and
$x^R$ using eq.\refb{eg5}. This gives:
\be \label{eg16}
\psi_\Xi (\xl, \xr) = \wt\NN^{26} 
\exp\Bigl(-{1\over 2}\xl\cdot K\cdot \xl -{1\over 2} \xr\cdot K \cdot
\xr 
- \xl\cdot L\cdot\xr  \Bigr)\, , 
\ee
where
\be \label{eg17}
K = A^{+T} V A^+ = A^{-T} V A^-, \qquad L = A^{+T} V A^-\, .
\ee
The equality of the two forms for $K$ follows
from eq.\refb{egtwist} and the relation $CSC=S$. The superscript
$T$ denotes transposition.

{}From this it is clear that in order that the sliver wave-function
factorizes in the sense described in eq.\refb{eg10}, the matrix $L$
defined above
must vanish. While we 
have not attempted  
an analytic proof of this, we have
checked using level truncation that this 
indeed appears to be the case. 
In particular, as the level is increased, 
the elements of the matrix $L$
become much smaller than typical elements of the matrix 
$K$.  If $L$ vanishes,
then the sliver indeed has the form given in eq.\refb{eg10} with
\be \label{egfactor}
f(\xl)= \wt\NN^{13} 
\exp(-{1\over 2}\, 
\xl \cdot K \cdot\xl) \, .
\ee
In this form we also see that the functional $f$ is actually real.
This is expected since the sliver is twist even,
and it must then correspond to a
symmetric matrix in half-string space.

\subsection{Building orthogonal projectors} \label{smds22}

\bigskip
Given that the sliver describes a projection operator into a one
dimensional subspace, the following question arises naturally :
is it possible to construct a projection operator into an orthogonal one
dimensional subspace? If we can construct such a projection operator
$\chi$, then we shall have 
\be \label{eg18}
\Xi * \chi = \chi * \Xi = 0, \qquad \chi * \chi = \chi\, ,
\ee
and $\Xi + \chi$ will satisfy the equation of motion \refb{eg1} and
represent a configuration of two D-25-branes.

{}From eq.\refb{eg10} it is clear how to construct such an orthogonal
projection operator. We simply need to choose a
function $g$ 
satisfying the same normalization condition 
\refb{normc} as
$f$  
and orthonormal to $f$: 
\ben \label{eg19}
&& \int [\hbox{d}y] \, f^*(y)
g(y) = 0\,,  \\
&& \int 
[\hbox{d}y]\, g^*(y) g(y) = \int 
[\hbox{d}y] \, f^*(y)f(y) \,  = 1 \,, \nonumber
\een
and then define
\be \label{egchi}
\psi_\chi(\xl,
\xr) = g(\xl) g^*(\xr)\, .  
\ee
There are many ways to construct such a function $g$, 
but one simple class of
such functions is obtained by choosing:
\be \label{eg20}
g(x^L) = \lambda_{\mu n} x^{L\mu}_n \, f(x^L)\, \equiv 
\lambda\cdot
x^L \, f(\xl)\,,
\ee
where $\lambda$ is a constant vector.
Since $f(-x)= f(x)$, the function $g(x)$  is orthogonal
to $f(x)$. For convenience we shall choose $\lambda$
to be real. Making use of \refb{egfactor} the normalization condition
\refb{eg19} for $g$ requires:
\be \label{eg21}
{1\over 2}  \lambda \cdot K^{-1} \cdot  \lambda
= 1\, .  
\ee
Additional orthogonal projectors are readily obtained.
Given another function
$h(x^L)$ of the form
\be \label{eg22}
h(x^L) = \lambda'\cdot x^L \, f(x^L)\, ,
\ee
with real $\lambda'$, 
we find another projector orthogonal to the sliver and to
$\chi$ if
\be \label{eg23}  
 \lambda\cdot K^{-1} \cdot \lambda' = 0, \qquad
{1\over 2} \,\lambda'\cdot K^{-1}  \lambda' = 1. \qquad
\ee
Since we can choose infinite set of mutually orthonormal vectors of this
kind, we can construct infinite number of projection operators into
mutually orthogonal subspaces, each of dimension one. By superposing $N$
of these projection operators we get a solution describing $N$ D-branes.

Note that these projectors are not manifestly Lorentz invariant
although, as we shall discuss in \ref{smdx}, they may well
be invariant under a combined Lorentz and gauge transformation, which
is all that is necessary.
On the other hand, we could also construct manifestly
Lorentz invariant projectors by taking $g(x^L)$ to be of the form $C_{mn}
x^{L\mu}_m x^{L\nu}_n  
\eta_{\mu\nu} + D$  
with appropriately chosen constants $C_{mn}$ and $D$.

One could continuously interpolate between the sliver $\Xi$ and the state
$\chi$ defined above via a family of rank one projection operators
$\chi_\theta$
characterized by an angle $\theta$. This corresponds to
choosing\footnote{Note that we could have used a general $U(2)$ rotation.
This will give rise to a complex $g(x^L)$.}
\be \label{ega1}
g_\theta (\xl) = (\sin\theta \, \lambda\cdot x^L + \cos\theta)
\, f(\xl)\, .
\ee

It is instructive to re-express the string state $\chi$, given by
eqs.\refb{egchi} and \refb{eg20} in the harmonic oscillator basis. 
We have, 
\be \label{eg24}  
|\chi\rangle = (\lambda \cdot \hat x^{L})\, (\lambda\cdot
\hat x^{R})
|\Xi\rangle\, .
\ee
Using eqs.\refb{eg6} and \refb{eg12} we can rewrite this as
\be \label{eg25}  
|\chi\rangle = -{1\over 4}\, ( \lambda\cdot  \wt A^+ E  
(a - a^{\dagger}) )\, (\lambda \cdot \wt A^- E  
(a -  a^{\dagger})) |\Xi\rangle\, .
\ee
We would like to express this in terms of only creation operators acting
on the sliver state $|\Xi\rangle$. For this we use the identity
\be \label{eg26}
a \,|\Xi\rangle = - S a^{\dagger}\,|\Xi\rangle \, .
\ee
Using this we can rewrite eq.\refb{eg25} as
\be \label{eg27}
|\chi\rangle = \Big( - \xi\cdot a^{\dagger} \,\, \wt\xi
\cdot a^{\dagger} + \kappa\Big) |\Xi\rangle\, ,
\ee
where $\wt\xi_{\mu n} = 
\xi_{\mu m}C_{mn} $, and  
\ben \label{eg28} 
\xi &=& {1\over 2}\, \lambda \cdot \wt A^+ E 
(1 + S)\, ,  \\
\kappa &=& {1\over 4 } \, \lambda\cdot \wt
A^+ E (1+S) E \wt A^{-T}\cdot \lambda
=  \xi \cdot (1+S)^{-1} \cdot \wt\xi\, . \nonumber
\een

The interpolating state $\chi_\theta$ can also be expressed in terms of
oscillators following an identical procedure. The result is
\be \label{ega2}
|\chi_\theta\rangle = \Big( - \sin^2\theta \, \xi\cdot a^{\dagger}\,
\, \wt\xi\cdot
a^{\dagger} - i \sin\theta \cos\theta (\xi + \wt\xi) \cdot
a^{\dagger} +  \cos^2 \theta +
\kappa \sin^2\theta \Big)
|\Xi\rangle\, .
\ee

Before we conclude this section, 
we would like to emphasize that
neither the sliver nor the state $\chi$ can be thought of as a projector
into a one dimensional subspace of the {\em full string Hilbert space}
even at zero momentum.
If we think of the zero momentum string field as a matrix, then 
the full string Hilbert space is the space of all matrices. The sliver
(or $\chi$) is 
 represented by a projection
operator of the form $P_{ij}=u_i u_j$, and 
acting on a general matrix
$M_{ij}$, gives $(PM)_{ik} = u_i u_j M_{jk}\equiv 
u_i v_k$. The subset   
$S$ of
matrices of
the form $u_i v_j$ for fixed vector $u$ but arbitrary vector $v$ is closed
under matrix multiplication; furthermore multiplication by any element of
$S$ from
the left takes any matrix to this subset $S$. Thus this subset of matrices
can be thought of as an ideal of the algebra associated with the full
string Hilbert space, and the operator $P_{ij}$ projects
any element of the algebra into
this ideal. In fact in this case the set of matrices in $S$ is also
closed under addition, and hence $S$ can also be regarded as a subspace of
the full string
Hilbert space.

\subsection{Lower dimensional D-branes} \label{lowerd}

\newcommand{\xm}{x^M}

The above method is not directly applicable to the construction of lower
dimensional D-brane solutions since the string wave-function now has
additional dependence on the zero mode coordinates 
$x_0^\alpha$, 
or
equivalently the string mid-point coordinates 
$\{X^\alpha(\pi/2)\}\equiv 
\xm$
where $X^\alpha$ 
denotes directions transverse to the D-brane. The
string field is now a function of $x^L$, $x^R$ and $\xm$, and the star
product of two string fields $\psi$ and $\chi$ is given by
\be \label{emid1}
(\psi*\chi) \,(\xl, \xr; \xm) = 
\int 
[\hbox{d} y]\,\,
\psi(\xl,  y; \xm) \,\chi ( y, \xr; \xm)\, .
\ee
Note that 
on the right hand side of eq.\refb{emid1} the integration is
carried out over the modes of the half string only {\it without the
mid-point}.
In analogy with the discussion in the previous section we could try to get
a solution of the equation $\psi_P * \psi_P = \psi_P$ by taking $\psi_P$
of the
form:
\be \label{emid2}
\psi_P(\xl,
\xr) = f(\xl;\xm) f^*(\xr;\xm)\, ,
\ee
with
\be \label{emid3}
f(\xl;\xm) = \NN' \exp\big( - K'_{mn} (\xl_m - c_m \xm) (\xl_n - c_n \xm)
\big)\, .
\ee
Here $K'_{mn}$ is some appropriate matrix, $c_n$ is 
an
appropriate
vector,
and $\NN'$ is a suitable normalization constant such that
\be \label{emid4}
\int [\hbox{d}y]  f^*(y; \xm)  f(y; \xm) = 1 \,.
\ee
In writing down the above formul\ae\ we have suppressed the Lorentz
indices, but they can be easily put back.
We expect that for a 
suitable choice of $K$ and $c_m$, the state $\psi_P$
defined through eqs.\refb{emid2} and \refb{emid3} coincides 
with the D-$p$-brane solution constructed in ref.\cite{0102112}. Following
the methods of previous subsections, we can easily construct other projection
operators orthogonal to $\psi_P$. We shall not pursue this line of
argument any further, however, 
and we now turn  
to the more rigorous algebraic formulation of the problem.

\sectiono{Multiple D-brane solutions $-$ Algebraic approach}
\label{smds}

In this section we give an oscillator construction of 
multiple brane solutions. While inspired by the considerations
of the previous section, the analysis is self-contained and
purely within the algebraic oscillator approach. Not only is the
algebraic approach less formal than the functional presentation
of the previous section, but it also
allows straightforward generalization
to the case of lower dimensional branes, as well as to the
case of non-coincident branes.

\subsection{Algebraic construction of higher rank projectors} \label{smds3}

Since the operators $\wt A^+$, $\wt A^-$ and $S$ are known explicitly,
eqs.\refb{eg27}, \refb{eg28} give an explicit expression for a string
state which
squares to itself and whose $*$-product with the sliver vanishes.
Since  the treatment of star products as delta functionals that glue
half strings in path integrals could conceivably be somewhat formal,
and also the  
demonstration that the sliver wave-functional factorizes was based on
numerical study,
in this section we will examine the problem algebraically using
the oscillator representation of star products. 
We shall give an explicit construction of the state
$|\chi\rangle$ without any reference to the matrices $\wt A^+, \wt
A^-$. For this we take a trial state of the same form as in  
eq.\refb{eg27}:
\be \label{eg29}
|\chi\rangle = \Big( - \xi\cdot a^{\dagger} \,\,\wt\xi\cdot
a^{\dagger} + \kappa\Big) |\Xi\rangle\, .
\ee
Here $\wt\xi \equiv C \xi$,  $\xi$ is taken to be an arbitrary
vector to be determined, and 
$\kappa$ is a constant to be determined. 
We shall actually constrain $\xi$ to satisfy
\be \label{eg30}
\rho_1 \xi = 0, \qquad \rho_2 \xi =\xi \, ,
\ee
where the $\rho_i$ are the projector operators
defined in \refb{proj}. 
We believe that $\xi$, as defined in eq.\refb{eg28}, automatically
satisfies eq.\refb{eg30} for any $\lambda$. This is the case if:
\be \label{exicons} 
\rho_1\, (1+S) \,E \,\wt A^{+T} = 0\, .
\ee
We do not have a proof of this equation, 
but numerical evidence  
indicates it is indeed true.  
At any rate, 
for the analysis
of this section we do
not use any of the results of the previous subsection. So we will
simply  proceed by taking $\xi$ to be an independent vector satisfying
eq.\refb{eg30}.  All the  results that follow 
are consistent with the results of the previous subsection if we assume
eq.\refb{exicons} to be correct.  

We first discuss some of the algebraic implications of
the constraints imposed in \refb{eg30}. Since $C\rho_1 C =
\rho_2$, eq.\refb{eg30} gives
\be \label{eg31}
\rho_2 \wt\xi = 0, \qquad \rho_1 \wt\xi =\wt\xi \, .
\ee
Since $\rho_1$ and $\rho_2$ are represented by symmetric
matrices, the transposed versions of  eqns. \refb{eg30} and
\refb{eg31} also hold.
Moreover, being projectors into complementary orthogonal
subspaces:  $\xi^T \wt \xi =
\xi^T (\rho_1 +\rho_2) \wt \xi = 0$. In fact, since $\rho_1$ and
$\rho_2$ commute with $M^{rs}$,
$X$ and $T$, as follows from eqs.\refb{ea222}, \refb{ea44a} and
\refb{proj}, we have the stronger
relation:
\be \label{ea11}
\xi^T f(M^{rs}, X, T) \, \wt \xi = 0\, ,
\ee
that holds for any function $f$.
Finally, we record the relations:
\ben \label{egiden}
M^{12} \xi &=& - X(1-T) \xi, \\
M^{21}\xi &=& (1 - TX) \xi\, . \nonumber
\een
The first follows directly from the first equation in \refb{ea8}.
The second follows from the first and the relation
$X+ M^{12}+ M^{21} = 1$.

\medskip
\noindent
We now begin the computation in earnest. This will have three
steps:

\smallskip
\noindent
(1) We require $\chi*\Xi =0$ and use this to fix $\kappa$.

\noindent
(2) We will see that indeed $\chi * \chi = \chi$ if $\xi$ is
 suitably normalized.

\noindent
(3) We will show that $\langle \chi | \chi \rangle = \langle \Xi
| \Xi \rangle$ and that $\langle \chi | \Xi\rangle=0$. 

\bigskip
We begin with step (1). The product $\chi*\Xi$ (and $\Xi * \chi$)
requires multiplying a sliver times a sliver with two oscillators
acting on it. As explained in section \ref{a2}, one can obtain this
result by
applying the differential operator
${\p^2\over\p
\beta_{1m\mu} \p \beta_{1n\nu}}$ on
both sides of the master equation eq.\refb{eb9}, and then setting
$\beta_{1}$ and
$\beta_{2}$ to zero. One finds
\be \label{eg32}
|\chi * \Xi\rangle = |\Xi * \chi \rangle = (\kappa + \xi^T
(\VV\KK^{-1})_{11}\wt\xi)|\Xi\rangle \, ,
\ee
where the matrix $\VV\KK^{-1}$, playing an important role
in this kind of calculation has been computed in
\refb{eb7}.
Since we require $\chi * \Xi = 0$, we need to set
\be \label{eg33}
\kappa = - \xi^T (\VV \KK^{-1})_{11} \wt \xi\,= -\xi^T T (1-T^2)^{-1} \xi\, ,
\ee
where the last equation in \refb{ehg} was used to simplify
the expression for $(\VV \KK^{-1})_{11}$.
On the other hand, using the relation $(1+S)^{-1} = (1-S) (1-S^2)^{-1} =
(1 - CT)(1-T^2)^{-1}$, and using eq.\refb{ea11} we can show that
eq.\refb{eg28} leads to precisely the same equation for $\kappa$. Thus we
see
that eq.\refb{eg33} agrees with the second equation in \refb{eg28}.
This concludes step one of the calculation.

\smallskip
In step two we 
calculate $\chi * \chi$ by differentiating both sides of
eq.\refb{eb9} with respect to $\beta_{1m\mu}$ and $\beta_{2m\mu}$
appropriate number of times, and then setting $\beta_1$ and $\beta_2$ to
zero. After using eq.\refb{ea11}, we get the
result for the $*$-product
to be:
\ben \label{eg34}
\chi * \chi &=& - (\xi^T (\VV \KK^{-1})_{12} \wt\xi)
\, \xi \cdot a^\dagger \,  \wt\xi \cdot
a^\dagger |\Xi\rangle  \\
&& + \Big( \big(\xi^T (\VV\KK^{-1})_{11} \wt \xi\big) \big( \xi^T
(\VV\KK^{-1})_{22} \wt \xi\big) + \big( \xi^T (\VV\KK^{-1})_{12} \wt
\xi\big) \big( \wt\xi^T
(\VV\KK^{-1})_{12} \xi\big) - \kappa^2\Big) |\Xi\rangle\, . 
\nonumber 
\een
Using \refb{eb7}, the last equation in \refb{ehg}, \refb{egiden},
and \refb{eg33}
one finds that  
\be \label{eg35}
\wt\xi^T
(\VV\KK^{-1})_{12} \xi = -\xi^T T (1-T^2)^{-1} \xi = \kappa\, .  
\ee
Furthermore $(\VV\KK^{-1})_{11} = (\VV\KK^{-1})_{22}$. Using this and
eqs.\refb{eg33}, \refb{eg35},  we
see
that eq.\refb{eg34} can be written as
\be \label{eg36}
\chi * \chi = (\xi^T (\VV\KK^{-1})_{12} \wt\xi) \Big( - \xi \cdot
a^\dagger \, \wt\xi \cdot a^\dagger
+ \kappa\Big)
|\Xi\rangle \,.
\ee
If we now normalize $\xi$ such that
\be \label{eg37}
\xi^T \, (\VV \KK^{-1})_{12}\, \wt\xi = 1\, ,  
\ee
then eq.\refb{eg36} reduces to the desired equation:
\be \label{eg38}
\chi * \chi = \chi\, .
\ee
The normalization condition eq.\refb{eg37} can be simplified using
eq.\refb{egiden} and the first equation in \refb{ehg} to 
obtain: 
\be \label{egkk1}
\xi^T (1 - T^2)^{-1} \xi = 1\, .
\ee
This concludes step two of the procedure.

\medskip
We shall now show that the new solution $\chi$ also represents  a single
D-25-brane. For this
we shall calculate the
tension associated with this solution and try to verify that it agrees
with the tension of the brane described by the sliver. Since the tension
of the brane associated to a given state is proportional to the BPZ norm
of the state \cite{0102112}, all we need to show is that
$\langle\chi|\chi\rangle$ is equal to $\langle\Xi|\Xi\rangle$. This is a
straightforward calculation using the result
\ben \label{eten1}
&& \langle 0| \exp\Big (-{1\over 2}  a\cdot  S a +
\lambda\cdot a \Big) 
\exp\Big (-{1\over 2} a\cdot S a^{\dagger} + 
\beta \cdot  a\Big) 
|0\rangle  \\
&=& \det(1 - S^2)^{-1} \exp\Big(\beta^T \cdot (1-S^2)^{-1} \cdot \lambda
-{1\over 2} \beta^T \cdot S (1-S^2)^{-1} \cdot \beta
\nonumber \\
&& \qquad \qquad -{1\over 2} \lambda^T \cdot S (1-S^2)^{-1} \cdot
\lambda\Big)\, ,\nonumber
\een
which follows from the more general overlap \refb{key}. One then
differentiates both sides of this equation with respect to components
of $\lambda$ and
$\beta$ to calculate the required correlator. The final result, after
using eqs.\refb{eg33},
\refb{egkk1}, is
\be \label{eten2}
\langle\chi | \chi\rangle = \langle \Xi|\Xi\rangle\, .
\ee
Thus the solution described by $\chi$ 
has the same tension as the solution
described by $\Xi$. Similar calculation also yields:
\be \label{einner}
\langle \Xi | \chi\rangle = 0\, .
\ee
Thus the BPZ norm of $|\Xi\rangle + |\chi\rangle$ is $2\langle\Xi |
\Xi\rangle$. This shows that $|\Xi\rangle + |\chi\rangle$ represents a
configuration with twice the tension of a single D-25-brane. This 
concludes the third and last step of the main calculation.\footnote{ 
The computations of BPZ inner products could have been done also by
using $\langle \chi | \chi \rangle = 
\langle \II | \chi\rangle$, as discussed
in subsection \ref{trid}.}

\bigskip
The new projector $\chi$ is not unrelated to the sliver
$\Xi$. One can 
prove directly that
$\chi$ is obtained from
$\Xi$ by a  rotation in the $*$-algebra. For this let us consider the
interpolating state given in
eq.\refb{ega2}:
\be \label{ega2new}
|\chi_\theta\rangle = \Big( - \sin^2\theta \, \xi \cdot a^\dagger \, 
\wt \xi \cdot  a^\dagger 
- i \sin\theta \cos\theta (\xi + \wt\xi) \cdot
a^\dagger  +  \cos^2 \theta +
\kappa \sin^2\theta \Big)
|\Xi\rangle\, ,
\ee
with $\xi$ now interpreted to be an arbitrary vector satisfying
eqs.\refb{eg30}, \refb{egkk1}, and $\kappa$ given by eq.\refb{eg33}. It
is straightforward to show using the techniques described earlier in this
subsection that
$\chi_\theta*\chi_\theta=\chi_\theta$. Furthermore, one can show that
\be \label{eggauge}
{d\chi_\theta\over d\theta} = (\chi_\theta * \Lambda - \Lambda *
\chi_\theta)\, ,
\ee
where
\be \label{eggauge2}
|\Lambda\rangle = i (\xi - \wt \xi). a^\dagger |\Xi\rangle\, .
\ee
Thus changing $\theta$ induces a rotation of $\chi_\theta$ by the
generator $\Lambda$. We shall call such transformations $*$-rotation. 
Since $\chi_{\pi/2}=\chi$ and $\chi_0=\Xi$, we get
\be \label{eggauge3}
\chi = e^{-\Lambda\pi/2} * \Xi * e^{\Lambda\pi/2}\, ,
\ee
where in the expansion of the exponential all products must be interpreted
as $*$ products. This establishes that $\chi$ is indeed a $*$-rotation
of the sliver. We would like to believe that 
this corresponds to a gauge
transformation in the full string field theory,  but we   
cannot settle this point without knowing the form of the kinetic
operator $\QQ$.
Nevertheless we have argued    
in refs.\cite{0102112,rsz} that under some
suitable assumptions 
$*$-rotations can indeed
be regarded as  gauge transformations.

\medskip
Consider now another projector $\chi'$ built just as $\chi$ but using
a vector $\xi'$:
\be \label{egp29}
|\chi'\rangle = \Big( - \xi'\cdot a^\dagger \, 
\wt\xi'\cdot a^\dagger + \kappa'\Big) |\Xi\rangle\, ,
\ee
with
\be \label{egp30}
\rho_1 \xi' = 0, \qquad \rho_2 \xi' =\xi' \, ,
\ee
$\kappa'$ given as 
\be \label{egpaa1}
\kappa' = -\xi^{\prime T} T (1-T^2)^{-1} \xi'\, ,
\ee
and normalization fixed by
\be \label{egp37}
\xi^{\prime T} (1 - T^2)^{-1}\xi' = 1\, .
\ee
Thus $\chi'$ is a projector orthogonal to $\Xi$. We now want to find the
condition under which $\chi'$ projects into a subspace orthogonal to
$\chi$
as well, {\it i.e.} the condition under which $\chi * \chi'$ vanishes. We
can compute $\chi * \chi'$ in a manner identical to the one used in
computing $\chi * \chi$ and find that it vanishes if:
\be \label{egpn1}
\xi^{T} (1-T^2)^{-1} \xi' = 0\, .
\ee
Since this equation is symmetric in $\xi$ and $\xi'$, it is clear that
$\chi'*\chi$ also vanishes when eq.\refb{egpn1} is satisfied.
Given eqs.\refb{egp37} and \refb{egpn1} we also have:
\be \label{egpn2}
\langle\chi'|\chi'\rangle = 1, \quad \langle\chi|\chi'\rangle=
\langle\Xi|\chi'\rangle = 0\, .
\ee
Thus $|\Xi\rangle+|\chi\rangle+|\chi'\rangle$ describes a solution with
three D-25-branes. 
This procedure can be continued indefinitely to generate solutions with
arbitrary number of D-25-branes.

\subsection{Lorentz invariance 
and Chan-Paton factors} 
\label{smdx} 

Although the sliver is a Lorentz invariant state, the state $\chi$ defined
in eq.\refb{eg29} is apparently not Lorentz invariant since it involves
the vector $\xi_{\mu m}$. A
general Lorentz transformation takes this state $|\chi\rangle$ to another
state
$|\chi'\rangle$ of the form:
\be \label{el1}
|\chi'\rangle = 
\Big( - \xi' \cdot a^{\dagger} \,\,\wt\xi'\cdot 
a^{\dagger} +
\kappa\Big)
|\Xi\rangle\, ,
\ee
where $\xi'_{\mu m}$ is the Lorentz transform of $\xi_{\mu m}$ and
$\wt\xi' = C\xi'$. Since $\rho_1$ commutes with a Lorentz transformation,
we have $\rho_1\xi'=0$. It is easy to see that the new
state is a $*$-rotation of the original state $\chi$. One way to show
this is to rotate $\chi'$ to $\Xi$ using the analog of
eq.\refb{eggauge3}   
with primed variables
and then rotate $\Xi$ back to $\chi$ using 
eq.\refb{eggauge3}. 
Thus although $\chi$ is not invariant under a Lorentz transformation by
itself, a combination of Lorentz transformation and a
$*$-rotation leaves the state
invariant.

A similar
analysis can be used to show that the multiple D-25-brane 
solution constructed earlier
is invariant under a Lorentz transformation  accompanied 
by a $*$-rotation. This $*$-rotation can be constructed according to
the following algorithm. 
Under a general Lorentz transformation the 
set of vectors $\xi^{(i)}$ used to construct the $N$ D-brane solution
gets rotated into another set of vectors $\xi^{\prime (i)}$.
The initial set of vectors $\xi^{(i)}$ as well as the final
set of vectors  $\xi^{\prime (i)}$ each define a set of $N$
orthonormal vectors in half-string state  space.
In general, given  two sets of $N$ orthonormal vectors in a vector
space there is a rotation taking one set to the other which can 
be written as the exponential of a generator. We can therefore
construct the generator that rotates the final set of $N$ vectors
into the initial set in the half string state space.
Given this generator we can explicitly represent
it as a state $\Lambda$ in the full string state space via the correspondance
between matrices in half-string space and string fields. This
$\Lambda$ is the generator of the desired $*$-rotation.

This, however, does not prove the Lorentz
invariance of these solutions since at present $*$-rotation in the matter
sector only appears to be a symmetry of the equations of motion for the
restricted class of field configurations of the form \refb{eo3}.
If, as discussed in
refs.\cite{0102112,rsz}, 
$*$-rotation can be regarded as a gauge
transformation
in the full string field theory, then the above analysis
would establish Lorentz invariance of our solutions.

A solution describing $N$ coincident D-25-branes corresponds to a
projection operator into an $N$-dimensional subspace of the half-string
state space.  If $*$-rotation really describes gauge transformation,
then the $U(N)$ symmetry acting on this $N$-dimensional subspace of the
half-string state space 
is a gauge symmetry. This can be
identified as the $U(N)$ gauge symmetry of $N$ coincident D-25-branes,
$-$ in
the same
spirit as in the description of D-branes as non-commutative
solitons \cite{0005031}. 

\subsection{Multiple D-$p$-brane solutions for $p<25$} \label{smds4}

In the previous 
subsections we have described methods for
constructing
space-time independent solutions of the matter part of the field equation
$\Psi * \Psi = \Psi$
which have vanishing $*$-product with the sliver. Taking the superposition
of such a solution and the sliver we get a solution representing two
D-branes. In this subsection we shall discuss similar construction for
the D-branes of lower dimension.

The explicit solution of the field equations representing D-$p$-branes of
all $p\le 25$ have been given in ref.\cite{0102112}. These solutions have
explicit dependence on the zero mode $x_0^\mu$ of the coordinate fields
transverse to the D-brane, and hence the method of subsection \ref{smds2}
relying on the factorization of space-time independent 
string functionals 
is not directly applicable.\footnote{The setup of subsection \ref{lowerd}
may apply.}
The algebraic
method of subsection
\ref{smds3}, however, is still applicable 
because
the solutions associated
with directions transverse to the D-brane have  a similar structure  to
the solutions associated with directions tangential to the D-brane. Thus,
for example, if $x^\bmu$
denote directions tangential to the D-$p$-brane ($0\le \mu\le p$) and
$x^\alpha$ denote
directions transverse to the D-brane ($p+1\le \alpha\le 25$), then the
solution representing the
D-$p$-brane has the form:
\be \label{esq1}
|\Xi_p\rangle = \NN^{p+1} \exp\Big(-{1\over 2}\eta_{\bmu\bnu} S_{mn}
a_m^{\bmu\dagger} a_n^{\bnu\dagger}\Big)|0\rangle
\otimes (\NN')^{25-p} \exp\Big(-{1\over 2} S'_{mn}
a_m^{\alpha\dagger} a_n^{\alpha\dagger}\Big)|\Omega\rangle\, ,
\ee
where in the second exponential the sums 
over $m$ and $n$ run from 0 to
$\infty$,
$\NN'$
is an appropriate normalization constant determined in ref.\cite{0102112},
and $S'$ is given by an
equation identical to the one for $S$ 
(see  eqs.\refb{ea1},
\refb{ea222}, \refb{ea44}, and \refb{ea44a}) 
with all matrices $M^{rs}$, $V^{rs}$, $X$, $C$ and $T$ replaced by the
corresponding primed matrices. The primed matrices carry indices running
from 0 to $\infty$ in contrast with the unprimed matrices whose indices
run from 1 to $\infty$. But otherwise the primed matrices satisfy the
same equations as the unprimed matrices. Indeed, all the equations in
section \ref{s3.2} are valid with unprimed matrices
replaced by primed matrices. In
particular we can define $\rho_1'$ and $\rho_2'$ in a manner analogous to
eq.\refb{proj}. The equations in section \ref{a2} also generalize
to the case when $|\Xi\rangle$ is 
replaced by $|\Xi_p\rangle$ and
$\beta\cdot a^\dagger$ is interpreted as
$(\beta_{n\bmu}a^{\bmu\dagger}_n+ \beta'_{n\alpha}a^{\alpha\dagger}_n)$.
We now choose 
vectors $\xi_{\bmu m}$, $\xi'_{\alpha m}$
such
that
\be \label{esq2}
\rho_1 \xi_{\bmu} = 0, \qquad \rho_1' \xi'_{\alpha} = 0\, .
\ee
We also define
\be \label{esq3}
\wt \xi_\bmu = C  \xi_\bmu, \qquad \wt \xi'_\alpha = C' \xi'_\alpha,
\qquad
\kappa' = -\xi^T T (1-T^2)^{-1} \xi - \xi^{\prime T} T'
(1-T^{\prime 2})^{-1} \xi'\, ,
\ee
and normalize $\xi$, $\xi'$ such that
\be \label{esq4}
\xi^T (1 - T^2)^{-1} \xi + \xi^{\prime T} 
(1-T^{\prime 2})^{-1} \xi' = 1\, .
\ee
In that case following the procedure identical to that of
subsection \ref{smds3} 
we can show that the state:
\be \label{esq5}
|\chi_p\rangle = \Big( - (\xi_{\bmu } \cdot a^{\bmu\dagger} +
\xi'_{\alpha }\cdot
a^{\alpha\dagger}) ( \wt\xi_{\bnu }\cdot
a^{\bnu\dagger} + \wt\xi'_{\alpha }\cdot a^{\alpha\dagger}) +
\kappa'\Big)
|\Xi_p\rangle\, ,
\ee
satisfies:  
\be \label{esq6}
\chi_p * \Xi_p = \Xi_p * \chi_p = 0\, ,
\ee
and 
\be \label{esq7}
\chi_p * \chi_p = \chi_p\, .
\ee
Thus $\chi_p+\Xi_p$ will describe a configuration with two D-$p$-branes.
This construction can  be generalized easily  
following the
procedure of subsection \ref{smds3} to multiple D-$p$-brane solutions.

\subsection{Parallel separated D-branes} \label{smdy}

At this stage it is natural to ask if we could superpose two (or
more) parallel D-$p$-brane
solutions which are separated in the transverse direction. For this we
first need to construct the single D-$p$-brane state which is translated
in the transverse direction along some vector $\vec s$ (say). This is
easily constructed by noting that the generator of translation in the
transverse direction is the momentum operator $\hat p_\alpha$ along the
$x^\alpha$ direction. This is related to $a_0^\alpha$,
$a_0^{\alpha\dagger}$ through the relations:
\be \label{epar1}
\wh p^\alpha = {1\over \sqrt b} (a_0^\alpha + a_0^{\alpha\dagger})\, ,
\ee
where $b$ is a constant introduced in ref.\cite{0102112}. Then the shifted
D-brane solution is given by
\be \label{epar2}
|\Xi^{(\vec s)}_p\rangle = \exp(i s_\alpha \wh p^\alpha) |\Xi_p\rangle\, .
\ee
Since translation is a symmetry of the action, this  state must
square to
itself. This is instructive to check explicitly. It suffices to work along
a single specific direction, and to focus on the portion of the
state in \refb{esq1} that involves that direction. This gives
\ben
\label{bkh}
&&\exp(i s \wh p) \exp\Big(\hskip-3pt-{1\over 2}
a^{\dagger}
\cdot S'\cdot a^{\dagger}\Big)|\Omega\rangle
\\
&& =  \exp (- {s^2\over 2b} ) \exp({i s\over \sqrt{b}} a_0^\dagger) 
\exp({i s\over \sqrt{b} }a_0) \exp\Big(\hskip-3pt -{1\over 2}
a^{\dagger}
\cdot S'\cdot a^{\dagger}\Big)|\Omega\rangle
\nonumber \\
&& =  \exp \Bigl(- {s^2\over 2b}(1-S'_{00}) \Bigr) 
\exp({i s\over \sqrt{b}}
(1-S')_{0m}a_m^{\dagger})   
\exp\Big(\hskip-3pt -{1\over 2}
a^{\dagger}
\cdot S'\cdot a^{\dagger}\Big)|\Omega\rangle \,, \nonumber
\een
where use was made of the Baker-Campbell-Hausdorf relation.
We now claim that the final state is actually of the form of the states 
$P_\beta \equiv \exp \Bigl(  
{\cal C} (\beta, \beta) \Bigr)  |\Xi_\beta\rangle$ 
introduced in section \ref{a2}
(see \refb{pbeta}). One identifies 
\be \label{ebetam}
\beta_m = -{is\over \sqrt{b}} ((1-S')C')_{0m}=  
-{is\over \sqrt{b}} (1-T')_{0m} \,,
\ee
where in the last step we have used 
$S' = C'T'=T'C'$ to recognize  
that $S'_{0m} = T'_{0m}$. A similar replacement can be done
in the first two exponentials of \refb{bkh}. It is now a simple
computation to verify that the first exponential
on the right hand side of \refb{bkh}  
correctly arises as $\exp (\CC (\beta,\beta))$ as required by
the identification. Since $P_\beta * P_\beta = P_\beta$ this
gives the desired confirmation.

We note, however, that the $*$-product of 
$\Xi^{(\vec s)}_p$ 
with $\Xi_p$ does not vanish manifestly.\footnote{
Although $\Xi^{(\vec s)}_p * \Xi_p$ is not manifestly
zero, we believe that on general grounds
\cite{rsz} such products do vanish. In this case, this
could happen as a result of the vanishing of $\exp (- \CC (0,\beta))$ for
the particular $\beta$ in equation \refb{ebetam}. This
is discussed further in \cite{rsz}.}
Hence we do not attempt to represent two
displaced D-branes as  $\Xi_p + \Xi^{(\vec s)}_p$. 
Let us consider, however, a
different projector 
built as in \refb{esq5}, with $\xi_{\bar \mu m} =0$ but $\xi^{\prime\alpha}_m \not= 0$:
\ben 
 \label{edif1}
&&|\chi_p\rangle = (-\xi' \cdot a^\dagger\, \wt\xi'\cdot a^\dagger +\kappa')
|\Xi_p\rangle, \qquad \wt\xi' = C' \xi',\qquad \rho_1'\xi'=0\,, 
\nonumber \\   \cr 
&& \kappa' =
-\xi^{\prime T} T' (1-T^{\prime 2})^{-1} \xi'\, ,\qquad
\xi^{\prime T}  (1-T^{\prime 2})^{-1} \xi' = 1\,. 
\een
If we
define $\eta'=\rho_2'\beta$, $\wt\eta'=\rho_1'\beta$, 
with $\beta$
given in eq.\refb{ebetam}, then the relation $C'\beta = \beta$, which
follows from eq.\refb{ebetam} leads to $\wt\eta' = C'\eta'$, as
the notation suggests.  It is then easy to
verify that\footnote{This
can be done by differentiating 
$\Xi_{\beta_1}*\Xi_{\beta_2}$ with respect to components of $\beta_1$ or
$\beta_2$.}
 \be \label{edif2}
\Xi^{(\vec s)}_p * \chi_p =  \chi_p * \Xi^{(\vec s)}_p = 0\, ,
\ee
if
\be \label{edif3}
\xi^{\prime T} (1 - T^{\prime 2})^{-1} \eta' =   0\, . 
\ee
Thus in this case $\chi_p + \Xi^{(\vec s)}_p$ is a solution of  
the equations of motion, and describes parallel separated D-branes.

We shall see in the next subsection how a simple procedure 
allows us to obtain  more general D-brane configurations.

\subsection{Intersecting D-branes} 
\label{smds5}

In this subsection we shall discuss the construction of multiple D-brane
solutions of different dimensions 
at different positions.

It is clear from our earlier discussion that in order to construct
multiple brane solutions, we need to construct states $\Psi_1$, $\Psi_2$,
$\ldots$ $\Psi_n$ satisfying 
\be \label{ess1}
\Psi_i * \Psi_j = \delta_{ij} \Psi_i\, ,
\ee
and then simply consider the solution 
\be \label{ess2}
\Psi = \sum_i \Psi_i\, .
\ee
Thus the question is: in
general how do we construct solutions $\Psi_i$ representing D-branes of
different types and still satisfying eq.\refb{ess1}?

In order to carry out this construction we need to assume that all the
D-branes that we want to superpose have one tangential (or transverse)
direction in common. For definiteness we shall take this to be the time
direction $x^0$. We shall now take the $\Psi_i$'s to be states of
the factorized form:
\be \label{ess3}
|\Psi_i\rangle = |\chi^{(i)}\rangle_0 \otimes |\psi^{(i)}
\rangle_{space}\, .
\ee
Here $|~\rangle_0$ denotes that we are describing a state in the Hilbert
space of the BCFT associated with $X^0$ and $|~\rangle_{space}$ denotes
that we
are describing a state in the Hilbert space of the BCFT associated with the
spatial coordinates. 
Since for such states the $*$-product factorizes into the $*$-product in
the time part and
the space part, we have
\be \label{ess5}
|\Psi_i * \Psi_j\rangle = |\chi^{(i)}*\chi^{(j)}\rangle_0 \otimes
|\psi^{(i)}*\psi^{(j)}\rangle_{space}\, .
\ee
The idea now is to choose the space part $|\psi^{(i)}\rangle_{space}$ to
be the space part of arbitrary
D-$p$-brane solutions 
(with possible shifts and different values of $p$)  
described earlier in this
section, or even the BCFT deformations discussed in ref.\cite{rsz}, but
to 
choose the $|\chi^{(i)}\rangle_0$ such that 
\be
\label{xzero} 
|\chi^{(i)} *
\chi^{(j)}\rangle_0 = \delta_{ij} |\chi^{(i)} \rangle_0\,.
\ee
This is achieved
by choosing  
\be \label{ess4}
|\chi^{(i)}\rangle_0 = -(-\xi^{(i)}\cdot a^{0\dagger}\,\,
\wt\xi^{(i)}\cdot
a^{0\dagger} + \kappa^{(i)} ) |\Xi\rangle_0,
\ee
with $\rho_1 \xi = 0$, and  
\be \label{ess7}
\kappa^{(i)} = -\eta_{00}\, \xi^{(i)T} T (1-T^2)^{-1} \xi^{(i)}, \qquad
\eta_{00}\, \xi^{(i)T}(1-T^2)^{-1} \xi^{(j)}=-\delta_{ij}\, .
\ee
The overall minus 
sign on the right hand side of eq.\refb{ess4} compared
to \refb{eg29} is due to
the choice of extra minus sign in the normalization condition given in the
second equation in \refb{ess7} 
as compared to \refb{egkk1}.  
This in turn is
required in order to find
solutions to the normalization condition.
This choice of $\Psi_i$ satisfies eq.\refb{ess1}, and hence $\Psi$ defined
in eq.\refb{ess2} satisfies the equation of motion. This has the
interpretation of superposition of D-branes of different
dimensions at different positions. 

\medskip
Before we end we  would like to note that not only could we have chosen
the $|\psi^{(i)}\rangle_{space}$ to be the space part of the solutions for
arbitrary D-$p$-branes at arbitrary positions, but with arbitrary
orientations as well. We simply take the space part of a known D-$p$-brane
solution and
apply a rotation to construct such a $|\psi^{(i)}\rangle_{space}$.
Superposition of states of the form $|\chi^{(i)}\rangle_0\otimes
|\psi^{(i)}\rangle_{space}$ will then give rise to configurations of
D-branes at different angles. A slight generalization of this allows us to
construct solutions describing superpositions of moving D-branes with
arbitrary velocities and arbitrary orientation, as long as all the
D-branes share one common space-like tangential coordinate, or one
common space-like
transverse coordinate which is orthogonal to the velocities of all the
branes. Let us
call this direction
$x^1$. We can then use this direction to construct a set of mutually
orthogonal projectors which will now play the role of
$|\chi^{(i)}\rangle$.\footnote{When $x^1$ is 
a common transverse coordinate orthogonal to the velocities 
and the branes are displaced
along it we use the construction at the end of section \ref{smdy} to
produce the required orthogonal factors.}
The $|\psi^{(i)}\rangle$'s are generated from the original D-brane solution
$\Xi_p$ by first removing the factor associated with the direction $x^1$,
and then applying combinations of translations, boosts and
rotations 
in directions perpendicular to $x^1$.
Superposition of solutions of the type 
$|\chi^{(i)}\rangle\otimes
|\psi^{(i)}\rangle$ will then give rise to a solution 
describing moving D-branes
with different velocities and orientations.
We believe that the restriction of having one common tangential or
transverse direction is only a technical difficulty, and a resolution
based on a different line of argument is presented in \cite{rsz}.

\section{Discussion}
\label{s10}

We now offer some brief remarks on our results and discuss some
of the open questions.

\begin{itemize}

\item
The first point we would like to make is that the present work
gives credence to the idea that half-string functionals do play
a role in open string theory. At least for zero momentum string fields,
as explained here, it is on the space of half-string functionals
that the sliver is a rank-one projection operator. We  do believe that the
left/right factorization of the sliver is an exact result, and it would be
interesting to find an explicit proof using the relevant matrices.  

\item 
We have seen that the sliver is a rank-one projector, and we have
learned how to construct higher rank projectors. The identity string
field, on the other hand is an infinite rank projector. Since rank-$N$
projectors are associated to configurations with $N$ D-branes, one would
be led to believe that the identity string field is a classical
solution of vacuum string field theory representing a background with
an infinite number of D-branes. While no doubt technical complications
related to normalization and regularization would be encountered in
discussing concretely such background, it
is interesting to note that a background with infinite number of
D-branes 
is natural for a general  K-theory analysis of 
D-brane states \cite{0007175}.

\item We have shown that various rank-one projectors are equivalent
under $*$-rotation;  a similarity transformation generated
by a state built with oscillators acting on the sliver. We would
expect $*$-rotations to be gauge symmetries of vacuum string field
theory, but this issue seems difficult to settle without explicit
knowledge of the kinetic operator $\QQ$ of vacuum string field theory.
For more discussion (but not a resolution) of this question, the
reader may consult \cite{rsz}.

\item In the study of $C^*$ algebras and von Neumann algebras projectors
play a central role in elucidating their structure. We may  be
optimistic that having finally found how to construct (some) projectors 
in the star algebra of open strings, a more concrete understanding of
the gauge
algebra of open string theory, perhaps based on operator
algebras\footnote{For some readable introductory comments on the
possible uses
of $C^*$ algebras in $K$-theory see \cite{0102076}, section 4.}, may
be possible to attain in the near future. This would be expected to have 
significant impact on our thinking about string theory.

\end{itemize}

Vacuum string field theory appears to be rather promising.
Multiple configurations of branes appear to be as simple
in this string field theory as they are in non-commutative
field theory. In turn, non-commutative solitons are in many
ways simpler than ordinary field theory solitons. All in all
we are in the surprising position where we realize that
in string field theory some non-perturbative physics -- such as 
that related to multiple D-brane configurations -- could be argued
to  emerge more simply than the analogous phenomena does in ordinary field
theory.

\medskip
\bigskip

\noindent{\bf Acknowledgements}:
We would like to thank F. Larsen, 
D. Gaiotto, J. Minahan, S. Minwalla, N. Moeller,
P. Mukhopadhyay, M.~Schnabl, S. Shatashvili, S. Shenker, 
W.~Taylor, E.~Verlinde and E. Witten for useful discussions.
The work of L.R. was supported in part
by Princeton University
``Dicke Fellowship'' and by NSF grant 9802484.
The work of A.S. was supported in part by NSF grant PHY99-07949.
The work of  B.Z. was supported in part
by DOE contract \#DE-FC02-94ER40818.

\end{document}